\documentclass[12pt]{article}

\usepackage{amsmath,amssymb,graphicx} 
\usepackage{epsfig}
\usepackage{pstricks}
\usepackage{booktabs}
\usepackage{multirow}
\usepackage[bookmarks]{hyperref}
\usepackage{subfig}
\usepackage{graphicx}

\newcommand{\beq}{\begin{eqnarray}}
\newcommand{\eeq}{\end{eqnarray}}

\newcommand{\centeron}[2]{{\setbox0=\hbox{#1}\setbox1=\hbox{#2}\ifdim
                                        
\wd1>\wd0\kern.5\wd1\kern-.5\wd0\fi
\copy0

\kern-.5\wd0\kern-.5\wd1\copy1\ifdim\wd0>\wd1
                                       \kern.5\wd0\kern-.5\wd1\fi}}
\newcommand{\ltap}{\>\centeron{\raise.35ex\hbox{$<$}}
                               {\lower.65ex\hbox{$\sim$}}\>}
\newcommand{\gtap}{\>\centeron{\raise.35ex\hbox{$>$}}
                               {\lower.65ex\hbox{$\sim$}}\>}

\newcommand\ZZ{\hbox{\zfont Z\kern-.4emZ}}
\font\zfont = cmss10 

\begin{document}
\begin{titlepage}
\begin{flushright}{\tt LYCEN 2011-05 }

{\tt KCL-PH-TH/2011-16}
\end{flushright}

\vskip.5cm
\begin{center}
{\huge \bf 
Superluminal neutrinos \\
in long baseline experiments \\

and SN1987a
}

\vskip.1cm
\end{center}
\vskip0.2cm

\begin{center}
{\bf
Giacomo Cacciapaglia$^{(a,b)}$, Aldo Deandrea$^{(a)}$, Luca Panizzi$^{(a)}$}
\end{center}
\vskip 8pt

\begin{center}
 {\it (a) Universit\'e de Lyon, F-69622 Lyon, France; Universit\'e Lyon 1, Villeurbanne;\\
CNRS/IN2P3, UMR5822, Institut de Physique Nucl\'eaire de Lyon\\
F-69622 Villeurbanne Cedex, France } \\
{\it (b) King's College London, Department of Physics, the Strand, London WC2R 2LS, UK} \\
\vskip 8pt
{\tt  g.cacciapaglia@ipnl.in2p3.fr,\\ deandrea@ipnl.in2p3.fr, \\panizzi@ipnl.in2p3.fr} 
\end{center}

\vglue 0.3truecm

\begin{abstract}
\vskip 3pt
\noindent
Precise tests of Lorentz invariance in neutrinos can be performed using long baseline experiments such as 
MINOS and OPERA or neutrinos from astrophysical sources. The MINOS collaboration reported a measurement of the muonic neutrino velocities that hints to super-luminal propagation, very recently confirmed at 6$\sigma$ by OPERA.
We consider a general parametrisation which 
goes beyond the usual linear or quadratic violation considered in quantum-gravitational models. We also propose a toy 
model showing why Lorentz violation can be specific to the neutrino sector and give rise to a generic energy behaviour 
$E^\alpha$, where $\alpha$ is not necessarily an integer number. Supernova bounds and the preferred MINOS and OPERA regions show a tension, due to the absence of shape distortion in the neutrino bunch in the far detector of MINOS. 
The energy independence of the effect has also been pointed out by the OPERA results.
\end{abstract}
\end{titlepage}

\newpage
\section{Introduction}
\label{sec:intro}
\setcounter{equation}{0}
\setcounter{footnote}{0}

The investigation of the properties of neutrinos has provided important discoveries in the past, such as 
oscillations with large mixing angles and mass structures. Neutrinos also play a special role
in theories and models beyond the standard model of particle physics. 
However, many properties of neutrinos still await experimental tests, like for example the value of the masses 
and the nature and existence of right-handed neutrinos. 

Apart from the continuous effort on the theoretical side, in particular motivated by ideas from quantum gravity, recent 
years have seen a renewed interest in experimental tests of Lorentz symmetry in order to probe the presence of 
new fundamental scales or unconventional space-time structures. Stringent bounds can be 
put on deviations from the standard Lorentz symmetry structure of space-time.  
The most stringent bounds come from particles like photons, electrons and nucleons (see for example 
\cite{Kostelecky:2008ts} for a list of bounds). Probes in the neutrino sector can in no way be competitive with such 
strong bounds. However there are at least a couple of good reasons to investigate these effects in the neutrino sector. 
The first is that as Lorentz violations (LV) are not described by a well established and unique fundamental theory it is 
not clear if their possible manifestations arise in a similar way in all particle sectors. 
As an example Lorentz violation can be present for particles without conserved internal quantum numbers as 
photons and Majorana neutrinos and absent for particles with electric charge \cite{Ellis:2003if}.
The second reason is that neutrinos often play a special role in theoretical models. In fact right handed neutrinos are 
the only particles in the standard model (SM) which are invariant under all the gauge symmetries of the theory: their nature 
and even existence are therefore not clear yet.

The most direct way to test LV in neutrinos is to measure their velocity , which should be equal to the speed of light due to the extreme smallness of their masses.
Such measurement has been performed at Fermilab long ago. 
Considering neutrinos with an average energy
of 80 GeV, a measure of the relative velocity of neutrinos with respect to muons gave~\cite{bound79}: 
$$|\beta_\nu - 1|<4\times10^{-5}\,,$$ where $\beta_\nu=v_\nu/c$ and we assume that muons travel at the speed of light.
More recently, the MINOS collaboration reported the measurement of the speed of neutrinos of energy around $3$ GeV 
using the precise time of flight measurement in the far detector.
They reported a shift with respect to the expected time of flight of~\cite{Adamson:2007zzb}
$$ \delta_t = -126\pm32\mbox{(stat)}\pm64\mbox{(sys)}~\mbox{ns} \quad 68\%~\mbox{C.L.}\,,$$
which corresponds to a neutrino velocity $$\beta_\nu -1 = (5.1\pm3.9)\times10^{-5}\,,$$ 
summing linearly the statistical and systematic uncertainties. 
This measurement agrees at less than 1.4$\sigma$ with the speed of light, therefore it does not provide a strong evidence in 
favour of Lorentz violating effects. However, if we take at face value the measurement, it suggests that neutrinos may 
propagate at velocities superior to the speed of light. 
This hint to super-luminal neutrino propagation motivated us in exploring the possible origin of such an effect.

The very recent OPERA results seem to confirm this hint. 
The OPERA collaboration reported that there is a deviation in the time of flight of neutrinos 
 which is consistent with the MINOS results, 
but in this case the precision of the measurement allows to establish super-luminal propagation of neutrinos at 6$\sigma$ level~\cite{seminarOPERA}:
$$\delta_t = - 60.7\pm6.9\mbox{(stat)}\pm7.4\mbox{(sys)}\;\mbox{ns} \quad 68\%\; \mbox{C.L.}$$
with a velocity
$$\beta_\nu - 1 = (2.48\pm0.28\mbox{(stat)}\pm0.30\mbox{(sys)}) \times 10^{-5} \quad 68\%\; \mbox{C.L.}$$

This is a very intriguing result 
because it is extremely challenging to explain this apparent Lorentz Violation in a consistent theoretical framework.
One possibility discussed in the literature is that neutrinos propagate in an extra dimensional space and therefore they 
can travel through shortcuts compared to photons and other standard model particles, which are bound to a lower dimensional brane 
world~\cite{Ammosov:2000kj}. In such a scenario, therefore, the super-luminal propagation is an effective result of the 
space--time structure and Lorentz invariance is recovered once the full extra dimensional space--time structure is 
taken into account \cite{Csaki:2000dm}. This possibility has been used in the past to conciliate different neutrino oscillation 
results \cite{Pas:2005rb}.

Long baseline experiments are not the only way to test Lorentz Violation in neutrinos. Neutrinos are also produced together 
with photons by astrophysical sources. In principle crossing data from sources of neutrino and gamma rays can allow to 
check time coincidence or delay.  
Core collapse supernovae are formidable sources of neutrinos as almost the total energy of the explosion is carried away by a burst of neutrinos. The handful of events measured from the supernova SN1987a provides a 
powerful tool to bound scenarios of modification of neutrino velocities due to the huge distance of the source of the 
neutrinos, which is in the Large Magellanic Cloud at 51 kiloparsec from Earth.
Any small effect would therefore be largely amplified by the long time of flight.
There are two main observations that can be used to bound Lorentz violating effects and which have been widely 
considered in the literature: the spread in the detection times of the neutrinos $\Delta t \sim 10$ sec, and the offset 
between the arrival of neutrinos and photons $\Delta t_{\nu \gamma}$. 
The former is relatively solid, and can provide the strongest bound on many modification of neutrino physics, like for 
instance the presence of keV mass sterile neutrinos from extra dimensions~\cite{Cacciapaglia:2002qr}.
The latter is very model dependent and uncertain, as the precise delay between the arrival of neutrinos and the arrival of 
the first light from the explosion is unknown.
Moreover, the mechanisms of neutrino and photon release from the supernova core are different and therefore there may be 
an offset at the source. Nevertheless, the huge distance spanned by the neutrinos and photons allows to pose 
competitive bounds, as we will see in the following.

In this work we will consider Lorentz violating effects entering as a modification of the speed of ultra-relativistic neutrinos.
We introduce a general parametrisation of the LV term as a power law of the neutrino energy $\sim E^\alpha$, 
where $\alpha$ is a generic non integer number.
While integers are naturally generated by a local operator, non integer values for $\alpha$ can be generated, for instance, by 
conformal neutrinos or neutrinos propagating in warped extra dimensions.
On more general grounds, a non integer power will allows us to be as model independent as possible.
A key observation is that the effect of an energy dependent modification of the velocity
for neutrinos will generate both a delay (or advance) 
in the time of flight and a distortion of the bunch shape of the neutrinos, if the spectrum is not monochromatic as it occurs 
in the case of MINOS and OPERA.
The effect on the bunch shape has not been considered before and it will lead to important consequences on the 
compatibility of supernova bounds and the preferred MINOS region.
To ease the tension, we will also consider other functional dependencies of the energy identifying a step function as the more promising possibility to accommodate both the surprising results from MINOS and OPERA and the supernova bounds.

The paper is organised as follows: in section \ref{sec:lor} we discuss the general form of the Lorentz violating term 
and its possible origin in the context of conformal neutrinos in warped extra dimensions; in section \ref{sec:sn} we 
present the bounds from supernova data on such a parametrisation; in section \ref{sec:long} we present the results of our simulation of the MINOS data and compare the preferred region to the supernova bounds; finally in section \ref{sec:model} we show some 
alternative energy dependencies compared to supernova bounds and  MINOS and OPERA data before concluding in section \ref{sec:concl}.

\section{Models of Lorentz Violation}
\label{sec:lor}
\setcounter{equation}{0}
\setcounter{footnote}{0}

Special relativity encodes Lorentz symmetries and in particular relates mass, energy and momentum in the well 
known form of the dispersion relation $E^2 -\vec{p}^2=m^2$ (in units where $c=1$, which we will follow in this section). 
From this dispersion relation one can estimate the effect 
of the neutrino mass on the velocity in the limit of large energy with respect to the mass:
\beq
v = \frac{\sqrt{\vec{p}^2}}{E}=\sqrt{1-\frac{m_\nu^2}{E^2}} \simeq 1 -\frac{m_\nu^2}{2E^2}\,. 
\eeq
The deviation from the speed of light is therefore negligible as suppressed by the very small neutrino mass compared 
to the neutrino energies we will consider: for instance, for a neutrino mass of $1$ eV and an energy of $10$ MeV, the deviation is at the level of one part in $10^{14}$.

\subsection{Parametrisation using the dispersion relations}

Typically  Lorentz violating effects are parametrised in the dispersion 
relation by allowing an extra dependence on the energy $E$ of the particle and a new mass scale $M \gg E$ at which 
Lorentz violating new physics appear:
\beq
E^2-\vec{p}^2\pm \frac{E^{\alpha+2}}{M^\alpha}=0 \label{eq:disp}
\eeq
where, typically, only the cases $\alpha =1,\; 2$ are considered~\cite{Ellis:2008fc}. Such an approach is quite popular in 
the literature and we shall follow it here, after slightly generalising the usual formula. 
The cases of integer $\alpha$ correspond to LV operators added to the neutrino Lagrangian and generated by some new 
physics at the scale $M$, in particular $\alpha = 1$ ($2$) corresponds to a dimension $5$ ($6$) operator.
In this paper, we generalise this formula to non integer exponents $\alpha$, thus allowing us to perform a more model 
independent analysis (such an effect may in fact not derive from a Lagrangian description).
Moreover, as discussed in the following, there are models in extra dimensions where such a dependence on the energy 
arises naturally.

Implicitly, assuming the existence of a dispersion relation close to those of wave mechanics corresponds to the 
assumption that LV effects are a small modification of the usual picture in which particles are described by propagating 
waves. This assumption has the important consequence that measurements of time shifts with respect to the prediction 
of special relativity are linked to energy, therefore the measurement of a time shift in a bunch of particles with a 
distribution in energy will also imply a modification of the bunch shape during its evolution. This is for example relevant
when discussing long baseline neutrino experiments, like MINOS, OPERA and T2K.
In such cases, a measurement of velocity can be correlated to a measurement of the bunch shape, as we will discuss 
later in the case of MINOS data.

In terms of an effect on the velocity of propagation one can parametrise
\beq
v=\sqrt{1-\frac{E^\alpha}{M^\alpha}} \simeq 1 \pm \frac{E^\alpha}{2M^\alpha}\; . \label{eq:vel}
\label{eq:alpha}
\eeq
The results in the rest of the paper only depend on this form of the velocity, and they are independent on the specific model that generates such an energy dependence in the velocity. The dispersion relation in Eq.~\ref{eq:disp}, therefore, is to be considered as a specific example.
As a simple test, one can measure the time of flight of neutrinos from the source to the far detector in long baseline 
neutrino experiments: a search for a $\delta_t$ with respect to the speed of light propagation can be performed. 
The MINOS collaboration has followed this strategy. However, the fact that the velocity depends on the energy, 
suggests that a more detailed study is in order when the energy spectrum is known with sufficient statistics 
and precision. The origin of the effect is that neutrinos with different energy will experience different time delay or advance; the effect, therefore, does not directly depend on the form of the dispersion relation but on the velocity.
In any energy dependent modification of the velocity, 
 a time shift is necessarily correlated to a shape distortion of the neutrino bunch.

\subsection{A toy model for non integer $\alpha$}
\label{subsec::toymodel}

As a motivation for the dispersion relation formula (\ref{eq:disp}) for neutrinos, we introduce a simple toy model that 
naturally generates non integer exponents. This behaviour is somewhat unusual, in the sense that the 
addition of a Lorentz violating local operator to the standard model Lagrangian would 
bring an integer number of energy factors -- related to the number of derivatives in the operator.

Neutrinos occupy a special seat in the standard model: the left-handed lepton doublets can be paired to the Higgs doublet 
to form an operator invariant under the gauge symmetries of the standard model.
This means that this operator can be coupled to a singlet fermionic operator and generate a mass for the neutrino.
The standard ways are to couple it to a fermion field, the right-handed neutrino, via a small Yukawa coupling, or to 
couple it to itself in a dimension 5 operator that generates a Majorana mass for the neutrino (thus violating lepton 
number conservation).
The latter can be obtained in the see-saw scenario as a result of the integration of a heavy right-handed neutrino.
However the nature of the neutrino mass term is still unknown due to the lack of direct and indirect tests.
Therefore, it may well be that the right-handed neutrino is not a simple fermionic field.

\begin{figure}[tb]
\begin{center}
\epsfig{file=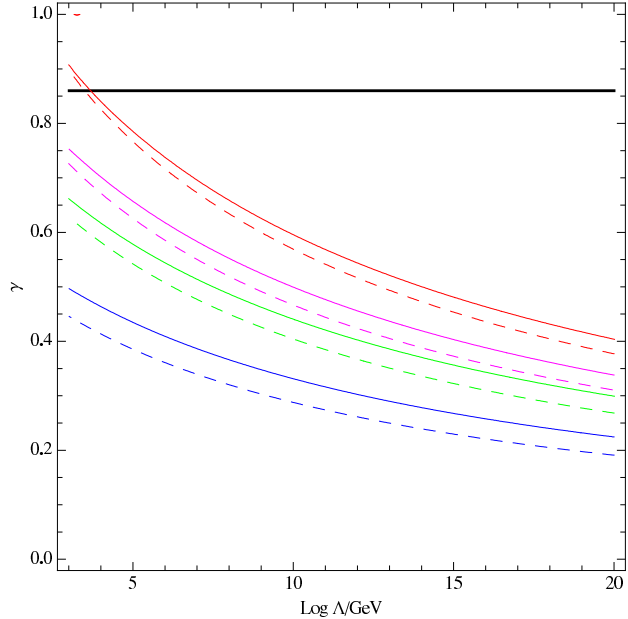, width=0.5\textwidth, angle=0}
\caption{\footnotesize Plot of the anomalous dimension as a function of the cut-off scale $\Lambda$ for different values of neutrino 
masses and Yukawa couplings: solid lines correspond to $m_\nu = 1$ eV while dashed lines to $m_\nu = \sqrt{\delta 
m^2_{solar}} = 0.05$ eV. The red lines correspond to $y_\nu = 1$, while the others to $y_\nu$ equal to the tau 
(magenta), muon (green) or electron (blue) Yukawa. The black horizontal line is the experimental bound 
$\lambda < 0.86$~\cite{vonGersdorff:2008is}.} \label{fig:models}
\end{center}
\end{figure}

One interesting possibility is that the right-handed neutrino is part of a conformally invariant sector of the theory
\cite{vonGersdorff:2008is,Grossman:2010iq}: in this paragraph we will summarise the results 
in~\cite{vonGersdorff:2008is} and 
formulate the same physics in terms of one extra dimension {\it \`a la AdS/CFT} (duality 
of a 4 dimensional strongly coupled conformal theory CFT to a weakly coupled 5 
dimensional anti-de Sitter space \cite{Maldacena:1997re}).
This is a natural expectation in the case of a sector that does not carry any of the quantum numbers of the standard model, in 
particular does not transform under gauge transformations.
Therefore, one of the bound states of the conformal sector, say $\psi_R$, can play the role of the right handed neutrino.
The main feature of a conformal operator is that it can have a large anomalous dimension $d_\psi = 3/2 + \gamma$, 
where $3/2$ is the canonical dimension of a fermionic field and $\gamma > 0$.
For $0 < \gamma < 1$, the dynamics of the operator $\psi_R$ can be described in terms of 
Unparticles~\cite{Georgi:2007ek}: the propagator can be written as
\beq
\Delta_\psi (p) = - i B_\gamma\, \frac{\bar{\sigma}^\mu p_\mu}{(-p^2 - i \epsilon)^{1-\gamma}}\,,
\eeq
where $B_\gamma = \frac{(4 \pi)^{- 2 \gamma} \Gamma (1-\gamma)}{\Gamma (1+\gamma)}$ is a normalisation factor 
which ensures that for $\gamma \to 0$ we obtain a standard fermionic propagator.
One can also rewrite the operator $\psi_R$ in terms of a canonically normalised field $\nu_R$
\beq
\psi_R = B_\gamma^{1/2} \mu^\gamma \nu_R\,,
\eeq
where $\mu$ is a renormalisation scale and the power $\gamma$ takes into account the anomalous dimension of the 
operator. The effective Lagrangian can now contain a Yukawa term between the Standard Model lepton doublet (which 
is an elementary field) and the CFT operator $\psi_R$:
\beq
\mathcal{L} = \frac{1}{\Lambda^\gamma} y_\nu \bar{L} H \psi_R + h.c. = B^{1/2}_\gamma 
\left( \frac{\mu}{\Lambda} \right)^\gamma y_\nu \bar{L} H \nu_R + h.c.
\eeq
Note that the Yukawa operator is irrelevant for $\gamma > 0$.
After the Higgs field develops a vacuum expectation value, this term will generate a mass term for the neutrinos
\beq
m_\nu = B^{1/2}_\gamma \left( \frac{m_\nu}{\Lambda} \right)^\gamma \frac{y_\nu v}{\sqrt{2}} \quad 
\Rightarrow \quad m_\nu = 
B_\gamma^\frac{1}{2(1-\gamma)} \left( \frac{y_\nu v}{\sqrt{2} \Lambda} \right)^\frac{\gamma}{1-\gamma} 
\frac{y_\nu v}{\sqrt{2}}\,. 
\label{eq:conformalneutrinomass}
\eeq
In the latter equations, we have fixed the renormalisation scale $\mu$ at the neutrino mass.
This formula offers an alternative to the see-saw mechanism with a Dirac mass for the 
neutrino~\cite{vonGersdorff:2008is}.
In Figure~\ref{fig:models} we show the correlation between the anomalous dimension $\gamma$ and the cut-off 
scale $\Lambda$ for different values of the neutrino mass and Yukawa coupling. From the plot we see that values 
of $0.2 < \gamma < 0.8$ are enough for cut-off energies up to the Planck mass.

The idea of conformal neutrinos can be elegantly reformulated in extra dimensions. 
In fact, a warped space~\cite{Randall:1999vf} with conformal metric
\beq
ds^2 =  \frac{R^2}{z^2} \left( dx_\mu dx^\mu - dz^2 \right)\,,
\eeq
describes a conformally invariant space: the rescaling of the co-ordinate $z$ compensates for the scaling of the 
co-ordinates of the 4 dimensional Minkowski space $x^\mu$.
A physical interpretation of the co-ordinate $z$ is to describe the red-shift of energy scales in the 4-dimensional world.
In order to build a viable model, we need to add a boundary for the space at small $z = \epsilon$.
Physically, the energy scale $\Lambda_{UV} = \frac{1}{\epsilon}$ corresponds to an ultra violet cut-off of the theory.
Fields can either live in the bulk of the extra dimension and depend on $z$, or they can be localised on the boundary at 
$z=\epsilon$. The AdS/CFT correspondence~\cite{Maldacena:1997re} offers us a way to physically interpret those 
fields: the fields living on the boundary correspond to elementary fields in the effective conformal theory, while fields 
living in the bulk correspond to operators of the CFT. Moreover, the gauge symmetries of the bulk are the same as the global symmetries of the 
conformal sector.

In order to reproduce the conformal neutrino scenario, we localise all the standard model fields, including gauge fields, on the UV 
boundary of the space $z = \epsilon$. The only field that is allowed to propagate in the bulk of the extra dimension is the right 
handed neutrino $\nu_R$, because it is a singlet under the gauge symmetries.
While the Lagrangian of the standard model fields is the usual 4D Lagrangian, for the right-handed neutrino one needs to write 
down a 5D Lagrangian which depends on the extra co-ordinate $z$.
After imposing the equations of motion on the bulk field, one can integrate the Lagrangian in $z$ and obtain an 
effective 4D Lagrangian. The neutrino sector of the model will then be described by the following Lagrangian:
\beq
\mathcal{L}_\nu = - i \bar{\nu}_L \bar{\sigma}^\mu \partial_\mu \nu_L + \frac{y_\nu v}{\sqrt{2}} (\nu_L \nu_R + h.c.) - i \Sigma(p) 
\nu_R \sigma^\mu \partial_\mu \bar{\nu}_R\,, \label{eq:effL}
\eeq
where~\cite{Cacciapaglia:2008ns}
\beq
\Sigma (p) = - \left( \frac{R}{\epsilon} \right)^2 \frac{\cos \alpha\, J_{c+1/2} (p \epsilon) + \sin \alpha\, J_{-c-1/2} (p \epsilon)}
{\cos \alpha\, J_{c-1/2} (p \epsilon) - \sin \alpha\, J_{-c+1/2} (p \epsilon)} \frac{1}{p \epsilon}\,,
\eeq
and $p = \sqrt{p^\mu p_\mu}$.
The angle $\alpha$ depends on the boundary conditions on the field for large $z$.
As we are interested in the physics at energies well below the cut-off of the theory, $\Lambda_{UV} = \frac{1}{\epsilon}$, 
we can expand $\Sigma$ for $p \epsilon \ll 1$: for $ c > - 1/2$ we obtain
\beq
\Sigma (p) \sim - \frac{R^4}{\epsilon^{3+2 c}} \frac{4^c \Gamma (1/2+c)}{\Gamma (1/2-c)} \tan \alpha \, p^{1-2 c} = N_c\,  
\left(\frac{p}{\Lambda} \right)^{1-2c}\,.
\eeq
From the effective Lagrangian in Eq.~(\ref{eq:effL}), we can calculate the propagator for the neutrino
\beq
\Delta_\nu \sim \frac{1}{\Sigma (p) p^2 - \left( \frac{y_\nu v}{\sqrt{2}}\right)^2}\,.
\eeq
The physical pole of this propagator defines the mass of the neutrino ($p^2 \to m_\nu^2$):
\beq
N_c \frac{m_\nu^{3-2c}}{\Lambda^{1-2c}} = \left( \frac{y_\nu v}{\sqrt{2}}\right)^2 \quad \Rightarrow \quad m_\nu = N_c^{\frac{1}
{3-2c}} \Lambda^{\frac{1-2c}{3-2c}} \left( \frac{y_\nu v}{\sqrt{2}} \right)^{\frac{2}{3-2c}}\,;
\eeq
therefore, comparing this formula with Eq.~(\ref{eq:conformalneutrinomass}), we identify $\gamma = 
c - 1/2$ and $N_c^{-1} = B_\gamma^2$.

In the extra dimensional model, Lorentz violation in the neutrino sector can be implemented in a very elegant way: in fact we can assume 
that the violation takes place by means of an operator in the bulk, while the physics on the UV boundary is Lorentz 
invariant. This would naturally explain why the other standard model particles do not feel the violation directly.
In the physical interpretation, it means that only the conformal sector violates Lorentz symmetry and neutrinos feel it 
because they have the most relevant coupling to the conformal operator~\footnote{Another relevant coupling involves a scalar operator coupled to the Higgs mass term~\cite{unhiggs}: this may induce large LV effects in the Higgs sector.}.
We will not discuss here the details of the Lorentz violating operator.
One simple way to model it is to assume that one of the sub-leading terms in the $p \epsilon$ expansion of 
$\Sigma$ only depends on the energy (and not on the invariant $p^2$):
\beq
\Sigma_{LV} (p) \sim  N_c\,  \left(\frac{p}{\Lambda} \right)^{- 2 \gamma} + \delta_{LV} \left(\frac{E}{\tilde{M}} \right)^{\beta} 
+ \dots
\eeq
where $\beta > - 2 \gamma$ for the expansion to be consistent.
In this case the propagator of the neutrinos will be modified and the Lorentz violating dispersion relation can be written as
\beq
p^2 + \frac{2 \delta_{LV}}{(1-\gamma) N_c} \frac{m_\nu^{2+2\gamma}}{\Lambda^{2 \gamma} \tilde{M}^\beta} E^\beta = m_\nu^2\,.
\eeq
The coefficient of the Lorentz violating term is suppressed by powers of the neutrino mass, however this is not a generic 
feature of these kinds of models but it depends on the particular choice of operator we made.
Here, we will take this as a hint of the possible existence of non integer exponents, and we will not pursue any further the 
construction of a specific model.

\section{SN1987a}
\label{sec:sn}
\setcounter{equation}{0}
\setcounter{footnote}{0}

In February 1987, a core collapse supernova, dubbed SN1987a, exploded in the Large Magellanic Cloud, about 51 
kiloparsec far from Earth. It is the closest supernova explosion recorded in recent times.
A few hours before the light from the supernova, a burst of neutrinos reached Earth and a handful of events have been 
measured by three neutrino detection experiments: Kamiokande II, IMB, and Baksan.
The neutrino burst lasted for about 10 seconds.
Even though the data are not very precise nor statistically rich, they are a powerful tool to pose bounds on various 
neutrino models, like for instance the presence of light sterile neutrinos or Lorentz violation in neutrino propagation.

\begin{table}
\centering\begin{tabular}{lccc}
\hline
    & $t_i$ & $E_i$& $\sigma_i$ \\
    \hline
Event  & (s)  &  (MeV) &  (MeV)  \\
\hline
\multicolumn{4}{c}{Baksan}\\
\hline
1  &$\equiv$ 0.0 & 12.0 & 2.4  \\
2  & 0.435 & 17.9 & 3.6  \\
3  & 1.710 & 23.5 & 4.7 \\
4  & 7.687 & 17.6 & 3.5 \\
5  & 9.099 & 20.3 & 4.1 \\
\hline
\multicolumn{4}{c}{IMB}\\
\hline
1  &$\equiv$ 0.0 & 38 & 7\\
2  & 0.412 & 37 & 7 \\
3  & 0.650 & 28 & 6 \\
4  & 1.141 & 39 & 7 \\
5  & 1.562 & 36 & 9 \\
6  & 2.684 & 36 & 6  \\
7  & 5.010 & 19 & 5 \\
8  & 5.582 & 22 & 5 \\
\hline
\end{tabular} \hspace{1cm}
\begin{tabular}{lccc}
\hline
    & $t_i$ & $E_i$& $\sigma_i$ \\
    \hline
    Event  & (s)  &  (MeV) &  (MeV)  \\
\hline
\multicolumn{4}{c}{Kamiokande II}\\
\hline
1 & $\equiv$ 0.0 & 20 & 2.9 \\
2 & 0.107          & 13.5 & 3.2  \\
3 & 0.303          & 7.5 & 2.0  \\
4 & 0.324 & 9.2 & 2.7  \\
5 & 0.507 & 12.8 & 2.9  \\
6 (omitted) & 0.686 & 6.3 & 1.7  \\
7 & 1.541 & 35.4 & 8.0  \\
8 & 1.728 & 21.0 & 4.2 \\
9 & 1.915 & 19.8 & 3.2  \\
10 & 9.219 & 8.6 & 2.7  \\
11 & 10.433 & 13.0 & 2.6  \\
12 & 12.439 & 8.9 & 1.9\\
13 (omitted) & 17.641 & 6.5 & 1.6  \\
14 (omitted)  & 20.257 & 5.4 & 1.4\\
15 (omitted) & 21.355 & 4.6 & 1.3 \\
16 (omitted) & 23.814 & 6.5 & 1.6\\
\end{tabular}
\caption{\footnotesize Data from SN1987a used for obtaining the limits on Lorentz violation. We have omitted 5 data points from 
Kamiokande II identified as a background events in previous investigations.
\label{table:datasn}}
\end{table}

The list of detection times, energies and corresponding errors on energies are given 
in Table \ref{table:datasn} for the data set of the three experiments. 
The uncertainties in the time measurements are in general much less than the statistical and 
energy uncertainties, and we therefore neglect them. Unfortunately, the relative arrival times in each experiment with 
respect to the others are not known, thus times are given setting $t\equiv0$ for the first event of every experiment, and 
the analysis must be performed independently for every data set. 
Moreover, we cannot fix the sign of the overall time shift of neutrino bunches with respect to the Lorentz conserving
hypothesis since the relative arrival time of the neutrinos with respect to light is known only with poor
accuracy \cite{Longo:1987gc,Stodolsky:1987vd}. We can therefore give the limits for either the super-luminal or the 
sub-luminal case but cannot distinguish the two. We can nevertheless compute the time shift in absolute value. 
Some of the data points for KII experiment are not included in the present analysis 
as identified as background events. On this point we have followed the results of 
\cite{Loredo:2001rx}: we have excluded events which fall below the energy threshold 7.5 MeV, which is known to 
be a large source of background. Since LV effects are energy dependent it is worth noticing that the energy measured 
at the detector is not the energy of the incoming neutrino, but that of the charged lepton resulting from the largely 
dominant absorption process \cite{Loredo:2001rx}:
\begin{equation}
 \bar\nu_e + p \to e^+ + n\,.
\end{equation}
The angular distribution of the emitted positron is to a good approximation isotropic and the energy of the incoming 
neutrino
is given by the relation $E_{\nu} = E_{l} + Q$, where $Q=1.29$~MeV is the neutron-proton mass deficit.
Gravity too influences the trip of the neutrinos from the supernova to the detector \cite{Longo:1987gc}, and small 
fluctuations in the gravitational field of the Galaxy can produce shifts in the arrival time of neutrinos. 
However, to perform this analysis we are assuming that gravitationally induced fluctuations in the time of flight of 
neutrinos are negligible.

Due to the assumption that Lorentz Violation is energy dependent, the time dispersion of the neutrinos observed in the 
detector may be in general different from the time dispersion at the supernova source. 
The time interval during which neutrinos are produced in a supernova is, however, model dependent, and various
scenarios have been studied in literature \cite{Loredo:2001rx}. 

In our analysis we will consider information coming from detected events and do not make any assumption on the 
production mechanism except for the functional structure of the energy spectrum at source \cite{Lunardini:2004bj} which 
is well established and anyway necessary to perform the calculation: 
\begin{equation}
F\sim E^{\alpha_z}\; e^{-(1+\alpha_z)E/E_0} 
\label{snspectrum}
\end{equation}
where $E_0$ and $\alpha_z$ are, respectively, the average energy of neutrinos and a pinching parameter.
The values of these parameters depend on details of the analysis and different techniques have been employed to 
determine them \cite{Loredo:2001rx,Lunardini:2004bj,Pagliaroli:2008ur}: we will give numerical results assuming 
$E_0=11$~MeV and $\alpha_z=3$, 
but we have checked that the dependence of results on these parameters is negligible in practice. 
 
 \begin{figure}[tb]
\centering
\subfloat[]
{\epsfig{file=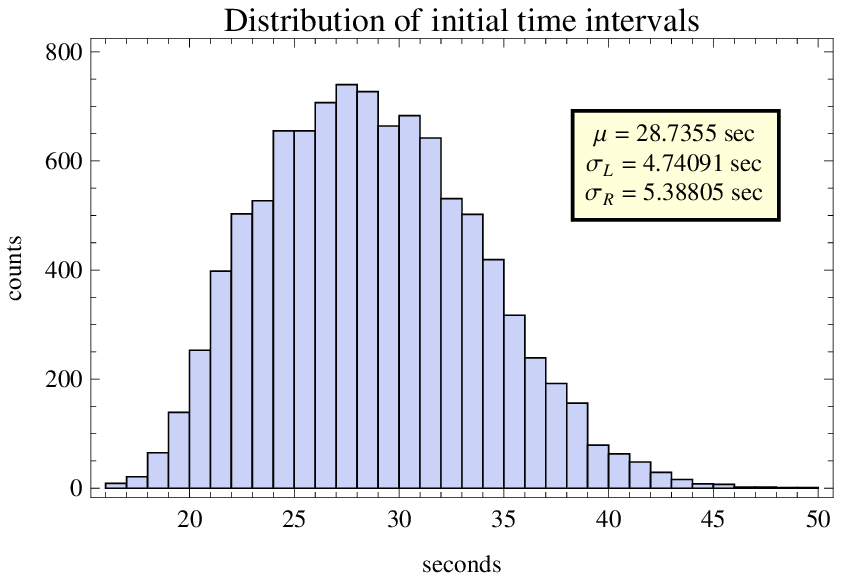, width=0.45\textwidth, angle=0}}
\subfloat[]
{\epsfig{file=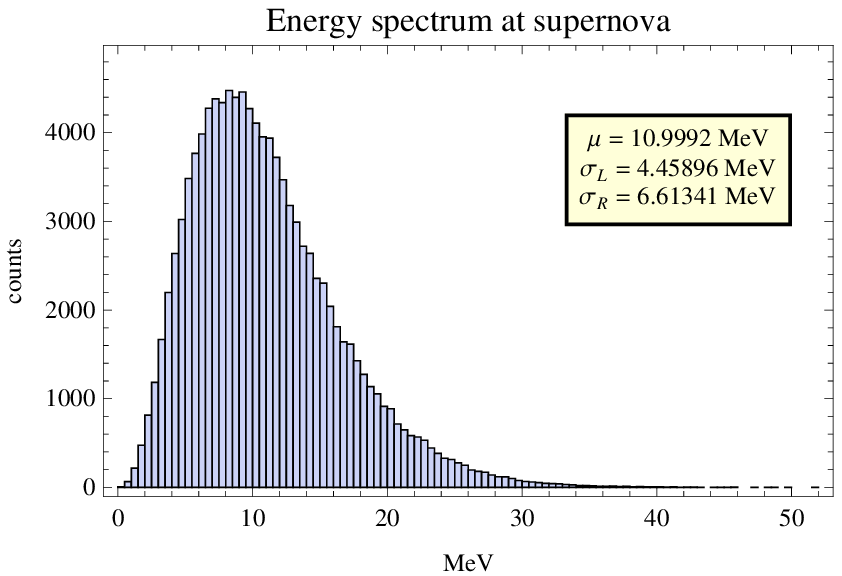, width=0.45\textwidth, angle=0}}\\
\subfloat[]
{\epsfig{file=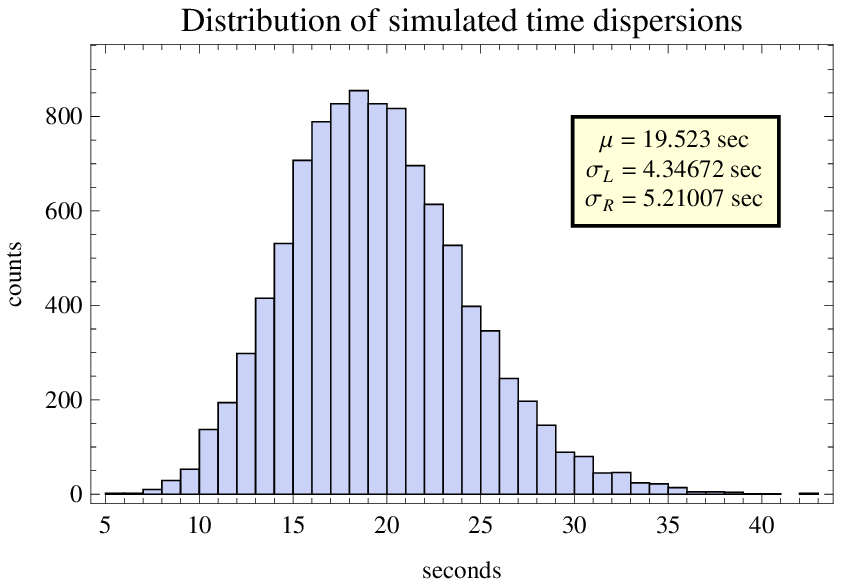, width=0.45\textwidth, angle=0}}
\subfloat[]
{\epsfig{file=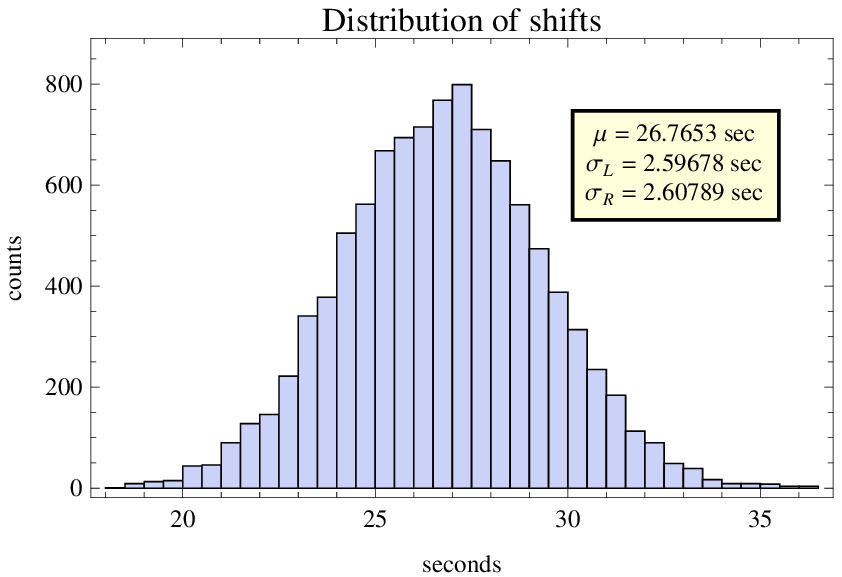, width=0.45\textwidth, angle=0}}
{\caption{\footnotesize Simulation steps (fixed number case) for KII data with LV parameters $\alpha = 3$, $E_0=11$~MeV and 
$M = 4\times10^9$ GeV.} 
\label{fig:examplesimKII}}
\end{figure}

Limits on LV parameters can be obtained simulating the evolution of a bunch of neutrinos from the supernova to the 
detector and measuring the probability that the time dispersion at detector predicted by LV parameters is consistent with 
the actually observed one within a given confidence interval. In order to be as model independent as possible and to 
keep the initial time dispersion of neutrinos at source as a free parameter, we evolved back the 
neutrinos observed in the three experiments including LV effects.
This calculation allows to estimate the production time for every parameter choice of the LV term, which become an input for the next step.
Afterwards, we simulated a neutrino burst of calculated duration and evolve it forward to the detectors, comparing the spread in 
arrival times with the measured one.
To take into account statistical errors and uncertainties in energy measurements, the simulation has been divided into the 
following steps:
\begin{enumerate}
\item we simulated $10^4$ neutrino sets at detector; for each set, the number of neutrinos and the detection 
times are the same as those measured, while the energies are randomly picked following a Gaussian distribution around the central value 
measured in the experiment with $\sigma$ given in Tab.\ref{table:datasn}. With this procedure, we have $10^4$ sets of neutrinos that correspond to the detected events.
\item every set has been evolved backward to the supernova source with fixed values of the LV parameters $\alpha$ and $M$. From each set, we can therefore calculate a time spread at the source.
\item we then simulated $10^4$ neutrino sets at the supernova source. Each set has the following properties, taking into account two possibilities for the analysis:
\begin{itemize}
\item[-] the number of neutrinos can be either the same as the observed ones (referred to as Fixed Number analysis in the following) or varying according to a Poisson distribution centred at 10 (Varying Number analysis):
in the latter case, the results are dependent on the choice of the expected number, 
however we have checked that varying this parameter in the interval \{8,12\} the results always remain within the same order of magnitude; 
\item[-] the duration of the neutrino burst is generated randomly following the distribution of initial time dispersions obtained in the previous step;
\item[-] the energies of neutrinos are distributed following the typical supernova spectrum (\ref{snspectrum}).
\end{itemize}
\item we finally evolved forward the simulated sets and obtained a distribution of time dispersions at the detector 
characterised by its average $\mu$ and (in general asymmetric) standard deviations $\sigma_L$ and $\sigma_R$. 
\end{enumerate}
Various distributions of a simulation with KII data have been summarised in Figure~\ref{fig:examplesimKII}. 
To obtain a bound on LV, for each choice of parameters we compare the observed time spread $\Delta t = 10$ sec with the simulated distribution: the values of parameters are excluded if $\Delta t$ falls outside of the interval $\{\mu-2\sigma_L,\mu+2\sigma_R\}$.

 \begin{table}[tb]
\centering\begin{tabular}{c|cc|cc|cc}
\toprule
\multirow{2}{*}{$\alpha$} & 
\multicolumn{2}{c|}{$\Delta t_{SN}$ (sec)} & 
\multicolumn{2}{c|}{$\Delta t_{\nu\gamma}$ (sec)} & 
\multicolumn{2}{c}{$M_{min}$ (GeV)} \\ 
\cmidrule{2-7} & FN & VN & FN & VN & FN & VN \\
\midrule
\multicolumn{7}{c}{Baksan}\\
\midrule
0.5      & $18.4^{+6.2}_{-5.6}$  & $11.7^{+3.3}_{-3.0}$ & $43.3^{+6.6}_{-7.0}$ & $22.2^{+3.1}_{-3.1}$ & $5 \times 10^{19}$ & $2 \times 10^{20}$ \\
1        & $18.0^{+6.5}_{-5.6}$  & $13.6^{+4.5}_{-4.0}$ & $22.1^{+5.0}_{-4.9}$ & $14.4^{+3.1}_{-3.0}$ & $2 \times 10^{9}$  & $3 \times 10^{9}$  \\
1.5      & $21.0^{+8.6}_{-6.9}$  & $14.5^{+5.4}_{-4.5}$ & $19.2^{+5.4}_{-4.9}$ & $11.4^{+2.9}_{-2.7}$ & $5 \times 10^{5}$  & $7 \times 10^{5}$  \\
2        & $22.3^{+10.7}_{-7.6}$ & $15.8^{+6.7}_{-5.2}$ & $16.9^{+5.1}_{-4.6}$ & $10.3^{+3.1}_{-2.8}$ & $8 \times 10^{3}$  & $1 \times 10^{4}$  \\
\midrule
\multicolumn{7}{c}{IMB}\\
\midrule
0.5      & $14.0^{+2.5}_{-2.1}$ & $15.1^{+2.7}_{-2.4}$ & $20.7^{+2.0}_{-2.2}$ & $22.2^{+2.3}_{-2.4}$ & $5 \times 10^{20}$ & $4 \times 10^{20}$ \\
1        & $16.8^{+3.0}_{-2.8}$ & $16.9^{+3.0}_{-2.7}$ & $16.5^{+1.9}_{-2.0}$ & $16.3^{+2.0}_{-2.0}$ & $6 \times 10^{9}$  & $6 \times 10^{9}$  \\
1.5      & $11.6^{+1.9}_{-1.6}$ & $11.6^{+1.9}_{-1.6}$ & $10.1^{+0.9}_{-1.0}$ & $10.0^{+1.0}_{-1.0}$ & $2 \times 10^{6}$  & $2 \times 10^{6}$  \\
2        & $16.8^{+3.7}_{-2.9}$ & $16.7^{+3.8}_{-3.0}$ & $13.1^{+1.6}_{-1.7}$ & $12.9^{+1.6}_{-1.6}$ & $2 \times 10^{4}$  & $2 \times 10^{4}$  \\
\midrule
\multicolumn{7}{c}{KII}\\
\midrule
0.5      & $30.4^{+4.5}_{-4.4}$ & $37.0^{+6.2}_{-5.9}$ & $40.6^{+3.4}_{-3.7}$ & $51.4^{+5.2}_{-5.3}$ & $1.6 \times 10^{20}$ & $9 \times 10^{19}$ \\
1        & $28.7^{+5.3}_{-4.7}$ & $34.8^{+7.1}_{-6.4}$ & $26.8^{+2.6}_{-2.6}$ & $32.7^{+4.2}_{-3.7}$ & $4 \times 10^{9}$    & $3 \times 10^{9}$  \\
1.5      & $27.3^{+6.4}_{-5.1}$ & $33.8^{+9.0}_{-7.2}$ & $21.7^{+2.3}_{-2.0}$ & $26.5^{+4.0}_{-3.1}$ & $1 \times 10^{6}$    & $8 \times 10^{5}$  \\
2        & $19.6^{+4.4}_{-3.1}$ & $19.7^{+4.5}_{-3.1}$ & $15.6^{+1.1}_{-0.8}$ & $15.8^{+1.4}_{-0.9}$ & $2 \times 10^{4}$    & $2 \times 10^{4}$  \\
\bottomrule
\end{tabular}
\caption{\footnotesize Average time dispersion of neutrino bunches at supernova ($\Delta t_{SN}$), average time shift between neutrinos and photons at detector ($\Delta t_{\nu\gamma}$) and lower bounds on the Lorentz Violating mass scale ($M_{min}$) for different exponent values ($\alpha$), in the Fixed Number (FN) and Variable Number (VN) hypothesis and for each experimental data set.
\label{boundsSN}}
\end{table}

The results obtained for different scenarios and for each experiment are shown in Table~\ref{boundsSN}. It is possible to see that our 
bounds for the mass scales are consistent with similar results obtained in other analyses \cite{Ellis:2008fc} and, noticeably, we 
obtain time scales for neutrino production in the supernova ($\sim$10 sec) which are consistent with models previously studied
in the literature \cite{Loredo:2001rx}.

So far we have only used the information about the time spread $\Delta t$ between neutrinos, which is sensitive to the 
energy dependence of the LV term.
The bound on the mass scale $M$ increases for small exponents $\alpha$, as shown in Table~\ref{boundsSN}, however 
this is due to the fact that the suppression from the energy dependence is milder.
For very small exponents, close to an energy independent modification of the velocity, we would expect this bound to 
disappear (as we will discuss later).
Another piece of information that can be used to pose bounds is the delay between the neutrinos and the photons.
This information is not very precise for two reasons: it is not very well known when the first light from the explosion 
reached Earth, and the emission times at source may well be uncorrelated due to the different emission mechanisms for 
neutrinos and photons.
Nevertheless, one can conservatively impose a bound of several hours on the delay $\Delta t_{\nu \gamma}$: 
following~\cite{Stodolsky:1987vd} we use 10 hours.
As shown by the values of $\Delta t_{\nu \gamma}$ in Table~\ref{boundsSN}, that we obtained with our simulation, we can 
see that this bound is not competitive with the bound from neutrino spread.

\section{Long baseline experiments}
\label{sec:long}
\setcounter{equation}{0}
\setcounter{footnote}{0}

Long baseline experiments, designed to study neutrino oscillations, have a unique capability to study the propagation of 
neutrinos if a precise time of flight measurement is possible.
At present, GPS based methods allow for sensitivities down to a few nanoseconds.
The long distance between source and far detector allows for a good sensitivity to LV effects.
Currently we can rely on the results by MINOS and OPERA collaborations.
The MINOS collaboration has published an analysis of the time of flight of neutrinos from Fermilab to the detector in the 
Soudan mine.
The result shows a deviation from what expected if neutrinos travelled at the speed of light, in particular neutrinos seem 
to arrive earlier that expected with a time shift of \cite{Adamson:2007zzb}:
\begin{equation}
\delta_t= -126 \pm 32 (stat.) \pm 64 (sys.) ~~\mathrm{ns}~~~~~ 68\%~~\mathrm{C.L.}
\end{equation}
which is consistent with the speed of light for the neutrinos at less than 1.4$\sigma$, but indicates a faster than light central 
value. The distance between the source of neutrinos and the far detector is $734 298.6 \pm 0.7$ m, which corresponds 
to a nominal time of flight $\tau = 2 449 356 \pm 2$ ns, while the mean neutrino energy is $\sim 3$ GeV.

OPERA, which is based in the Gran Sasso laboratory and utilises neutrinos from the CNGS beam at CERN, enjoys a 
similar distance between source and far detector, however neutrinos have a larger energy of about $20 \div 30$ GeV and 
a more precise time of flight measurement is possible with precision down to a nanosecond.
The results of the two experiments can therefore complement each other very effectively. However, due to the fact that 
the OPERA results were announced only very recently, we will present an analysis of MINOS data, keeping in mind that 
the same analysis can (and will) be performed once the data collected at OPERA will be available.

In the case of MINOS, there is more in the published data than just a time shift as the energy profiles are also
available. Therefore, we can use this information as a further constraint as the energy dependent LV dispersion gives in 
general both a shift in the arrival time and a distortion in the bunch structure. 
Moreover the two effects are correlated and not independent.
In order to perform our analysis, we extracted from the MINOS neutrino velocity measurement paper 
\cite{Adamson:2007zzb} the time distributions of neutrino events in the near and in the far detector. In the MINOS paper 
the time distribution of neutrinos observed in the far detector is shown superimposed to the expectation curves after 
having fitted the time of flight. The result of the fit is claimed to correspond to a shift of the plotted data points of +126 ns 
with respect to the measured data time distribution. We took into account this shift in order to obtain the original data 
time distribution. 

\begin{figure}[tb]
\begin{center}
\epsfig{file=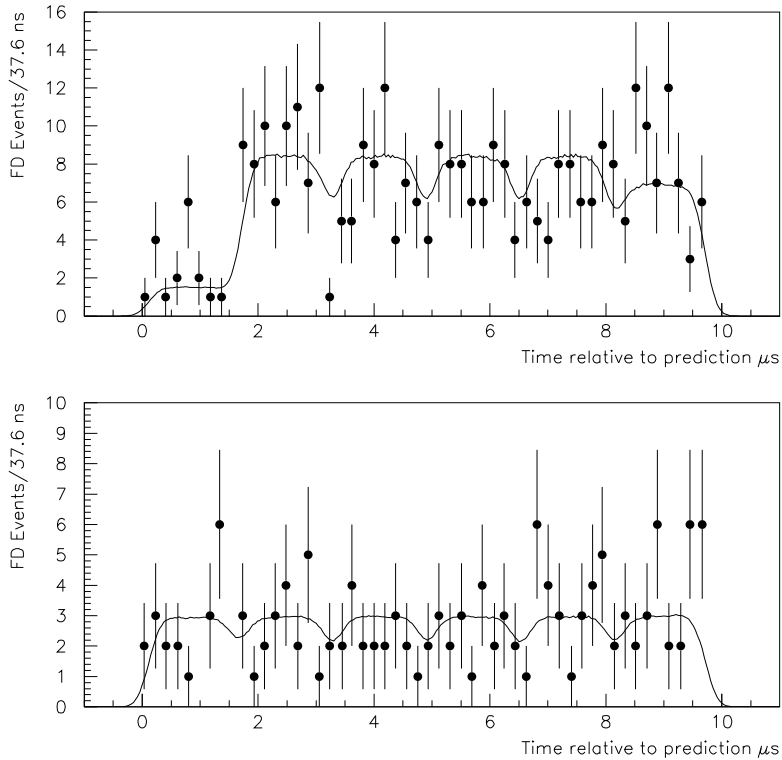, width=0.7\textwidth, angle=0}
\caption{\footnotesize Our expectation curves for the far detector MINOS data using as input the published 
near detector time distributions. We reproduce very well Fig.2 of the MINOS paper \cite{Adamson:2007zzb} 
for the far detector data points with a smearing of 150 ns.} 
\label{fig:minos}
\end{center}
\end{figure}

We computed our own expectation curves at the far detector for the 5 and 6 batch spills by using as input the published 
near detector time distributions and applying, as explained in the paper, a smearing of 150 ns describing the total relative 
far detector -- near detector time uncertainty. The expectation curves that we obtained, once superimposed to the far 
detector 
data points, reproduce very well Figure 2 of the MINOS paper \cite{Adamson:2007zzb}, as shown in Fig. \ref{fig:minos}.
In order to cross-check what we computed for the expectation curves and the data points at the far detector, we tried to reproduce 
the maximum likelihood analysis as it is described in the MINOS paper. Since the far detector data points we extracted from the 
MINOS paper were binned in time and more precise information on the event times was not available, in the likelihood calculation 
we randomly distributed the events in a uniform way inside each bin by preserving its normalisation. By maximising the likelihood 
function, computed on the basis of our expectation curves, we found a shift compatible with -126 ns, with a corresponding a 
statistical uncertainty of 32 ns. This result, similar to the published one, gave us confidence of being able to reproduce the MINOS 
data analysis. This was a mandatory condition in order to correctly further develop the analysis in the framework of LV models, 
which implies re-computing the expectation curves as a function of the $\alpha$ and $M$ parameters.

This analysis, although based on a similar principle, is more complex in terms of computing procedures than a maximum likelihood 
determination of a simple time shift. Given a pair of parameters ($\alpha$, $M$) the expectation curves can be 
obtained by taking into 
account the time distribution in the near detector, the smearing of 150 ns and the energy spectrum of events interacting in the far 
detector. The energy spectrum takes into account the neutrino oscillation disappearance effect on the charged current component. 
Given a bin in the near detector event time distribution, this is extrapolated to the far detector by performing a Monte Carlo 
simulation of a large sample of events, generated according to the spectrum of interacting neutrinos. For each event belonging to a 
given time bin in the near detector, the time at the far detector is computed by correcting for the LV shift as a function of 
($\alpha$, $M$)  and the neutrino energy and by including the Gaussian smearing of 150ns accounting for the time 
measurement uncertainties. In order to be compatible with the MINOS result in the application of our model 
we considered only negative time shifts.

\begin{figure}[tb]
\begin{center}
\epsfig{file=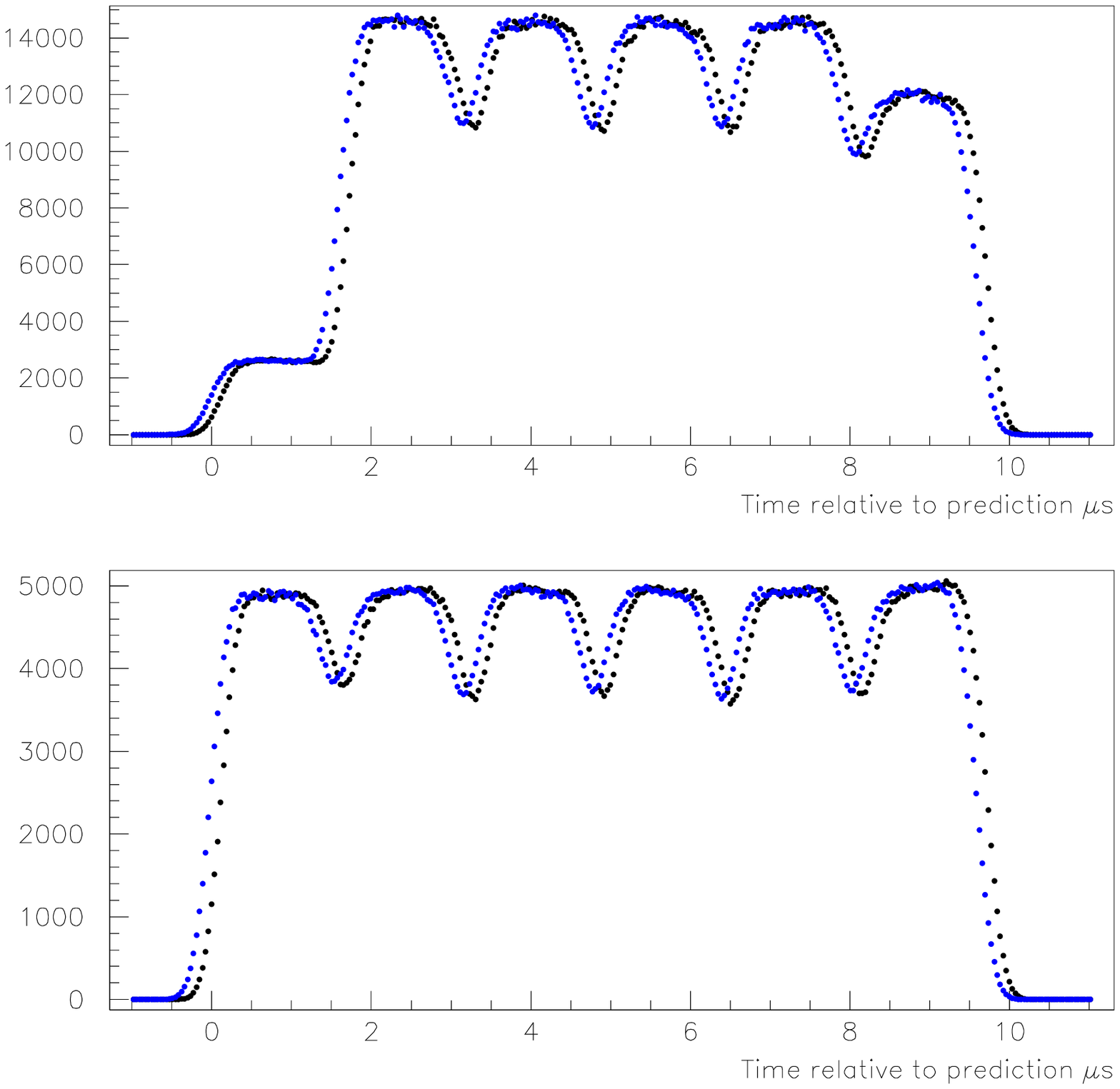,width=.49\textwidth}
\epsfig{file=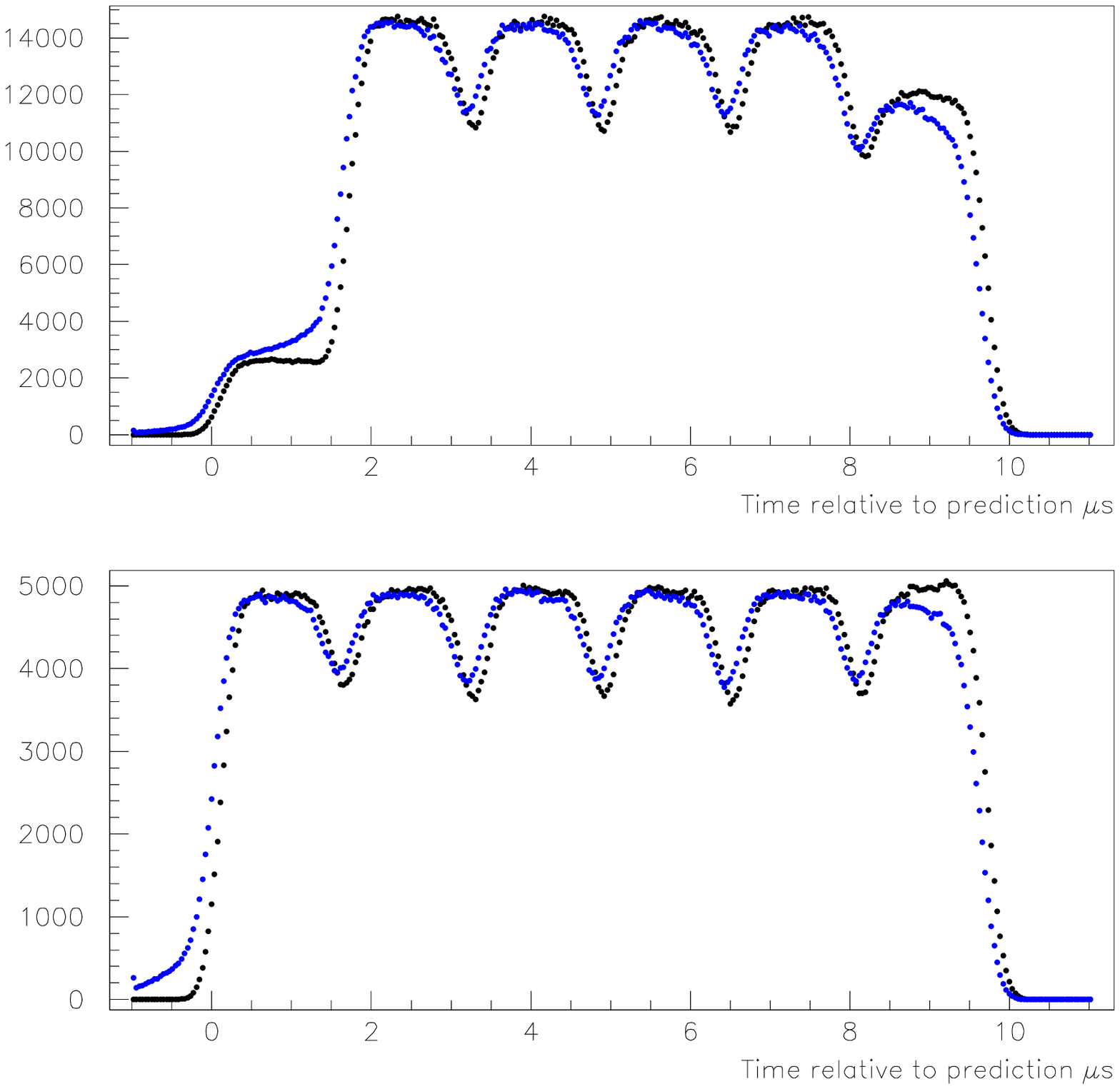,width=.49\textwidth}
\caption{\footnotesize Time distribution at the far detector is obtained by summing all the extrapolated contributions of the single 
bins in the near detector time distribution. The time distributions predicted at the far detector in 
absence of Lorentz violating effects (black curves) are superimposed to the curves (blue) foreseen by the LV model. Left 
panels for $\alpha=0.7$ and $M=2.31\times 10^6$ GeV (up--left 5 spills, down--left 6 spills), right panels for $\alpha=2.1$ 
and $M=429.9$ GeV  (up--right 5 spills, down--right 6 spills).} 
\label{fig:fardect}
\end{center}
\end{figure}

\begin{figure}[tb]
\begin{center}
\epsfig{file=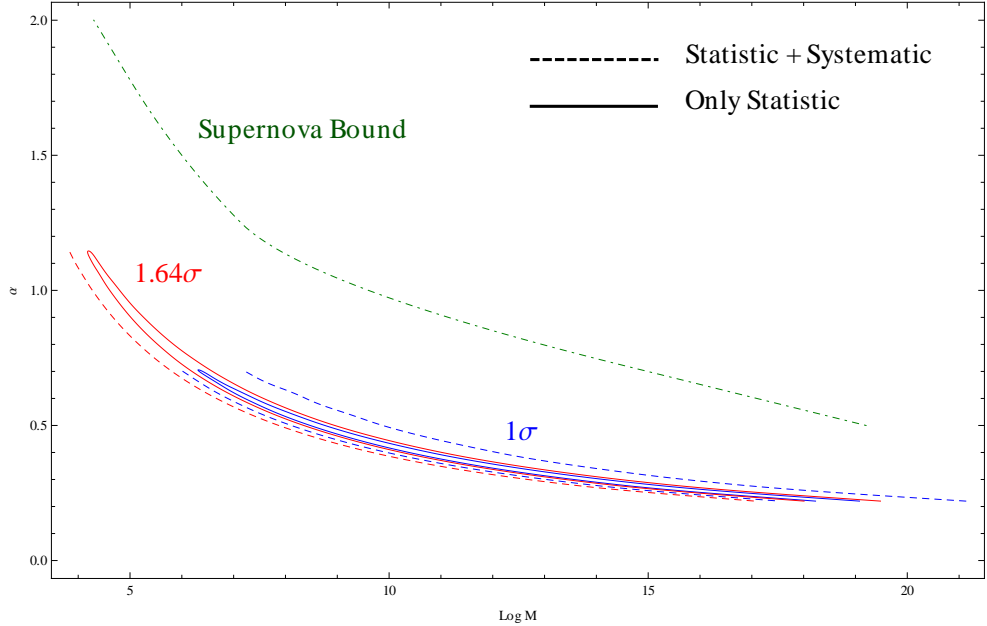, width=0.8\textwidth, angle=0}
\caption{\footnotesize Bounds coming from SN1987a combined with our fit of the MINOS data in the ($\alpha$, $M$) plane, 
which show tension between the MINOS neutrino velocity measurement and the SN1987a bound.} 
\label{fig:bound}
\end{center}
\end{figure}

The time distribution at the far detector is obtained by summing all the extrapolated contributions of the single bins in 
the near detector time distribution. Examples are shown in Fig. \ref{fig:fardect}.
We generated several sets of prediction curves in the ($\alpha$, $M$) plane. In particular, given a value of alpha, we 
performed a fine sampling as a function of M for values in the region expected to be interesting with respect to the effect 
measured by MINOS. We avoided values of M implying very large time shifts, by far not compatible with the MINOS 
measurement, or values well beyond the point where time shifts are unobservable. 
We computed the likelihood function for each point in the ($\alpha$, $M$) plane and  
parametrised its evolution as a function of M with a set of smooth curves. We performed an 
overall maximisation of the likelihood function in the ($\alpha$, $M$) plane and computed the contours of the 
allowed regions corresponding to different confidence levels. 
In our simulation, we did not take into account the systematic uncertainty of 64~ns affecting the  
measurement of the neutrino time of flight.
The main systematic uncertainty come from a limited knowledge of the length of cables in the electronics, therefore 
the main effect is to add an unknown shift to all the time measurements; however the measurement of the bunch shape 
should be marginally affected.
We have added and subtracted the 64ns from the results of our simulation, therefore enlarging the allowed 
range in $M$ for each given $\alpha$, as shown in Fig.\ref{fig:bound}.

The region which fits better the MINOS data corresponds to small values of $\alpha$. This is due to the fact that for such 
$\alpha$ values the distribution of time shifts is narrower and more similar to a global energy-independent time shift. 
For large values of $\alpha$, time shifts become more energy-dependent and the predicted time distribution at the far 
detector is not just a displaced replica of the near detector time distribution but its shape is changed as well. The 
distribution is distorted and affected by long time anticipation tails which are related to the tails in the neutrino energy 
spectrum. 

According to our analysis, and as shown in Fig.  \ref{fig:bound}, there is tension between the MINOS 
neutrino velocity measurement and the SN1987a bound. The MINOS measurement could become compatible with the 
SN1987a bound at high values of $\alpha$, which maximises the energy dependence of the time shift and large 
values would explain why such time shift is not observed with the SN1987a neutrinos which 
have energies by 3 orders of magnitude smaller than the MINOS neutrinos. However $\alpha$ maximises as well the 
effect of the energy spread in the neutrino spectrum. MINOS data are more compatible with a simple shift than with a 
energy dependent shift, and this points to the opposite direction, corresponding to low values of $\alpha$.

Considering MINOS data alone, the tension could be explained by a statistical fluctuation. In fact, the tension is 
completely removed at less than 2$\sigma$. The results from OPERA, however, point to the same direction and with 
much better precision, therefore we are lead to push further our analysis and try to understand the origin of the 
discrepancy between supernova and long baseline data.

One difference between Supernova and MINOS or OPERA data is the flavour of the neutrinos: supernova detectors 
only measured electron neutrinos, while the flavour in MINOS is muonic.
One may think of a flavour dependent LV effect which affects only the propagation of muon neutrinos. 
However, an effect of the size of the one measured by MINOS would completely destroy the neutrino oscillations: in fact, 
a different speed between components of different flavours in a neutrino propagating from the Sun to Earth would destroy 
the coherence between the two components much faster than the oscillation, thus inhibiting the oscillations 
\cite{Coherence}. In the following section we will take a different approach, that is to explore LV terms with different 
functions of energy.

\section{Alternative forms of Lorentz Violation}
\label{sec:model}
\setcounter{equation}{0}
\setcounter{footnote}{0}

The main result of our analysis is a tension between the bounds obtained from SN and MINOS data with a 
Lorentz violating power law term in the velocity for neutrinos.
The tension is mainly due to the energy dependence of the effect: for large values of the exponent $\alpha$ the 
supernova bound is loose due to the large suppression given by the small neutrino energies, however such region is 
disfavoured by MINOS due to the non observation of a distortion in the neutrino bunch at the far detector. This tension 
can only be worsened by OPERA data, due to their better precision.
One way to alleviate the tension is to modify the energy dependence of the LV term: in fact, supernova neutrinos have 
energies around 10 MeV, while MINOS uses neutrinos of $\sim 3$ GeV and higher for OPERA.
Therefore, supernova bounds and the results from long baseline experiments might be compatible if the energy dependence function in the LV term were sharper 
between the two energy regions.
In the following, we shall consider a generic LV parametrisation of the neutrino velocity, 
so that any energy dependent deviation from the usual Lorentz conserving velocity law can be written as:
\begin{equation}
 v = 1 + \Delta_{LV}(E)\,,
\end{equation}
where the sign is chosen to fit an advance, in accordance with MINOS data.
We considered three parametrisation:
\begin{itemize}
\item[-] a power law dependence
\beq
\Delta_{LV} = \delta \times (E/M_{Pl})^\alpha\,; \label{eq:powerlaw}
\eeq
note that this is an alternative parametrisation with respect to what we used in the previous section, the main difference 
being that the mass scale is kept fixed and equal to the Planck mass while a variable dimensionless coefficient $\delta$ is introduced.
The only reason for this is to be more sensitive to the region of low $\alpha$ (mild energy dependence), which we will focus on here.
\item[-] an exponential dependence
\begin{equation}
\Delta_{LV} = \delta \times \left(1-e^{-E/\mu}\right)\,, \label{eq:exponential}
\end{equation}
where the term becomes energy independent at large energies.
\item[-] a step function in terms of an hyperbolic tangent
\begin{equation}
\Delta_{LV} = \delta \times \left(1+\tanh\left({E-m'\over\mu}\right)\right)\,; \label{eq:step}
\end{equation}
in this case we can have a velocity close to 1 at low energies and an energy independent deviation at large energies.
\end{itemize}

\begin{figure}[t!]
\begin{center}
\subfloat[Power Law dependence. 
The allowed region is on the left of the blue and black lines.]
{\epsfig{file=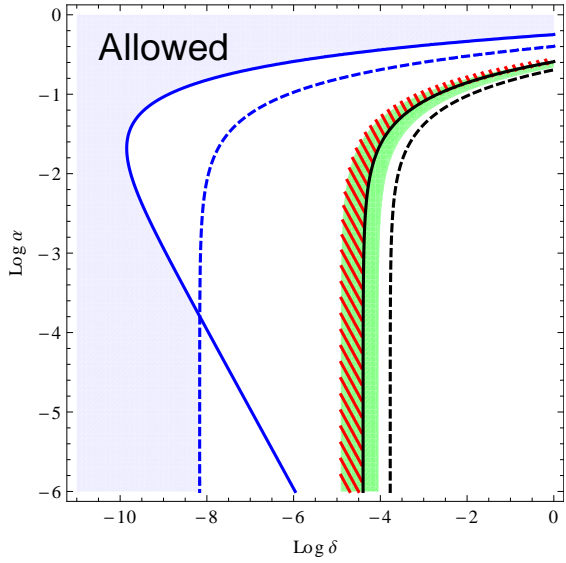, width=0.45\textwidth, angle=0}\label{fig:SN_MINOS_PL2}}\hfill
\subfloat[Exponential Dependence. 
The allowed regions are on the left of the blue and black lines.]
{\epsfig{file=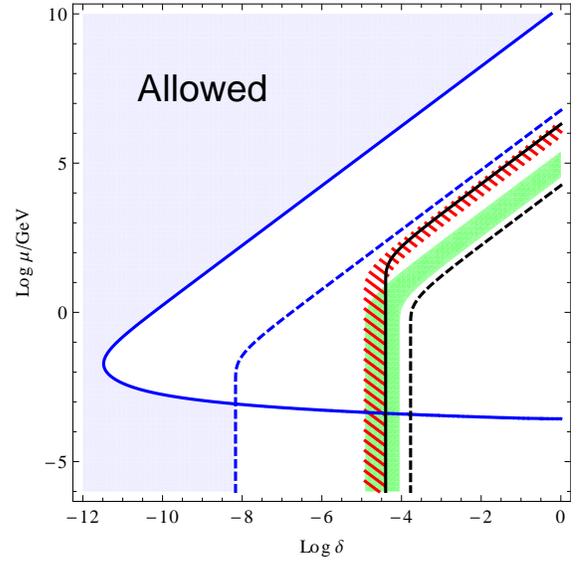, width=0.45\textwidth, angle=0}\label{fig:SN_MINOS_Exp}}\\
\subfloat[Hyperbolic tangent dependence with $m^\prime=1$ GeV.
The allowed regions are on the left of the blue and black lines.]
{\epsfig{file=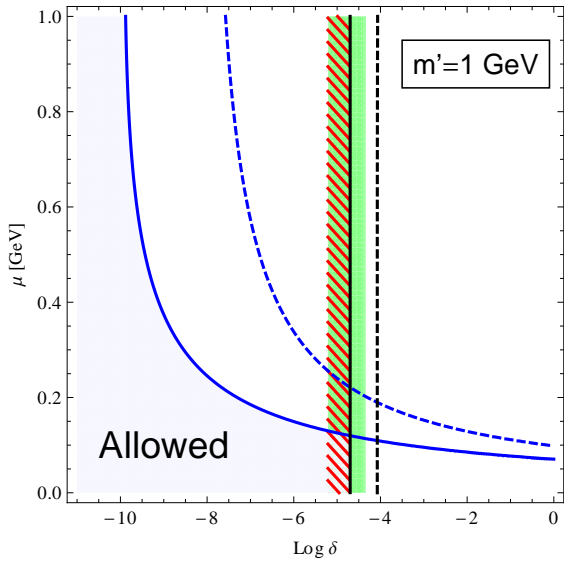, width=0.45\textwidth, angle=0}\label{fig:SN_MINOS_Tanh1}}\hfill
\subfloat[Hyperbolic tangent dependence with $m^\prime=4$ GeV. 
The allowed regions are on the left of the blue and black lines.]
{\epsfig{file=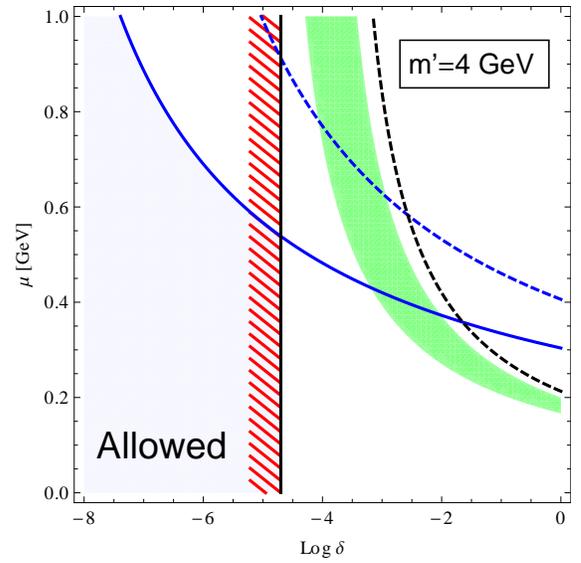, width=0.45\textwidth, angle=0}\label{fig:SN_MINOS_Tanh1a}}
\caption{\footnotesize Bounds for different parametrisation of the dispersion relation:
10 second spread between supernova neutrinos of different energy (solid blue), 
 10 hour offset between supernova neutrinos and photons (dashed blue),
high energy neutrino velocity at Fermilab (solid black), 
bound from MINOS data at 3$\sigma$ (dashed black). 
The green region is the preferred region from MINOS data at $1\sigma$, while the dashed red region
is the preferred region from OPERA at $3\sigma$.} 
\label{fig:ref}
\end{center}
\end{figure}

In this section we limit ourselves to an estimate of the bounds, by use of simple considerations, and we will not rely on a 
detailed simulation. The bounds we impose are:
\begin{itemize}
\item the Fermilab bound~\cite{bound79} on the velocity of high energy ($\sim$80 GeV) neutrinos obtained comparing the velocities of muons and neutrinos: 
$$|v_\nu -1| = |\Delta_{LV} (80\; \mbox{GeV})| <4\times10^{-5}\,.$$
\item the MINOS observed time delay $\delta_t$ at 1$\sigma$ and bound at 3$\sigma$: we have considered the time of flight of a neutrino with 
energy $E=3$ GeV (roughly at the peak of the spectrum) and imposed that $\delta_t$ must lie within 
$-126\pm(32+64)$ ns. 
The statistical and systematic errors quoted in Ref.\cite{Adamson:2007zzb}
have been summed linearly because the effect of the systematic error is an overall shift of the data; we have thus chosen the most conservative hypothesis, 
assuming that the shift induced by the systematic error is maximal and contributes in the same direction as the statistical error.
We have also considered the bound at 3$\sigma$, which is consistent with the Lorentz conserving hypothesis,
and excludes an advance of more than  $414$ ns. 
\item the offset between neutrinos and photons from SN1987a: as a conservative maximum interval we used 10 hours as estimated in \cite{Stodolsky:1987vd}. We considered
a neutrino with energy $E=40$ MeV, since this value is nearly at the higher end of the spectrum and, assuming that the LV effect increases with energy, neutrinos
with high energy would reach the detector before softer neutrinos. Moreover, a neutrino with an energy of $39$ MeV has been measured at IMB.
\item the spread in the arrival times of supernova neutrinos bounded below 10 seconds: we considered the propagation of two neutrinos at the opposite sides of the energy spectrum, namely $7$ and $40$ MeV. 
\end{itemize}

The results of our analysis are shown in Fig.~\ref{fig:ref}: in blue the bounds from supernova SN1987a from the offset between 
neutrinos and photons (dashed line) and the time spread in neutrinos (solid line); in solid black the bound from high energy neutrinos; 
in green the region preferred by MINOS at 1$\sigma$ (the dashed black lines represent the bound at 3$\sigma$); while the red dashed 
region is preferred by OPERA (at 3$\sigma$).

In the case of the power law (\ref{eq:powerlaw}), we checked that the rough estimate gives similar results as the detailed simulation 
presented in the previous section. For very small $\alpha$, the LV term becomes almost energy independent: the bound from the time 
spread in neutrino arrival times is loosened as expected because the velocities of neutrinos of different energy become very similar.
However, the bound from the offset with photons kicks in and shows still a tension: in fact, using the MINOS central value for the 
velocity, we would expect the neutrinos to reach Earth almost 9 years before the photons.
The same conclusion applies to the 3$\sigma$ OPERA preferred region, which overlaps to the 1$\sigma$ MINOS one in this case.

A similar behaviour also appears for the exponential dependence (\ref{eq:exponential}): for small $\mu$ the bound from the time spread is removed, however the offset with photons still shows a strong tension with the MINOS and OPERA preferred regions.

\begin{figure}[t!]
\begin{center}
\subfloat[]
{\epsfig{file=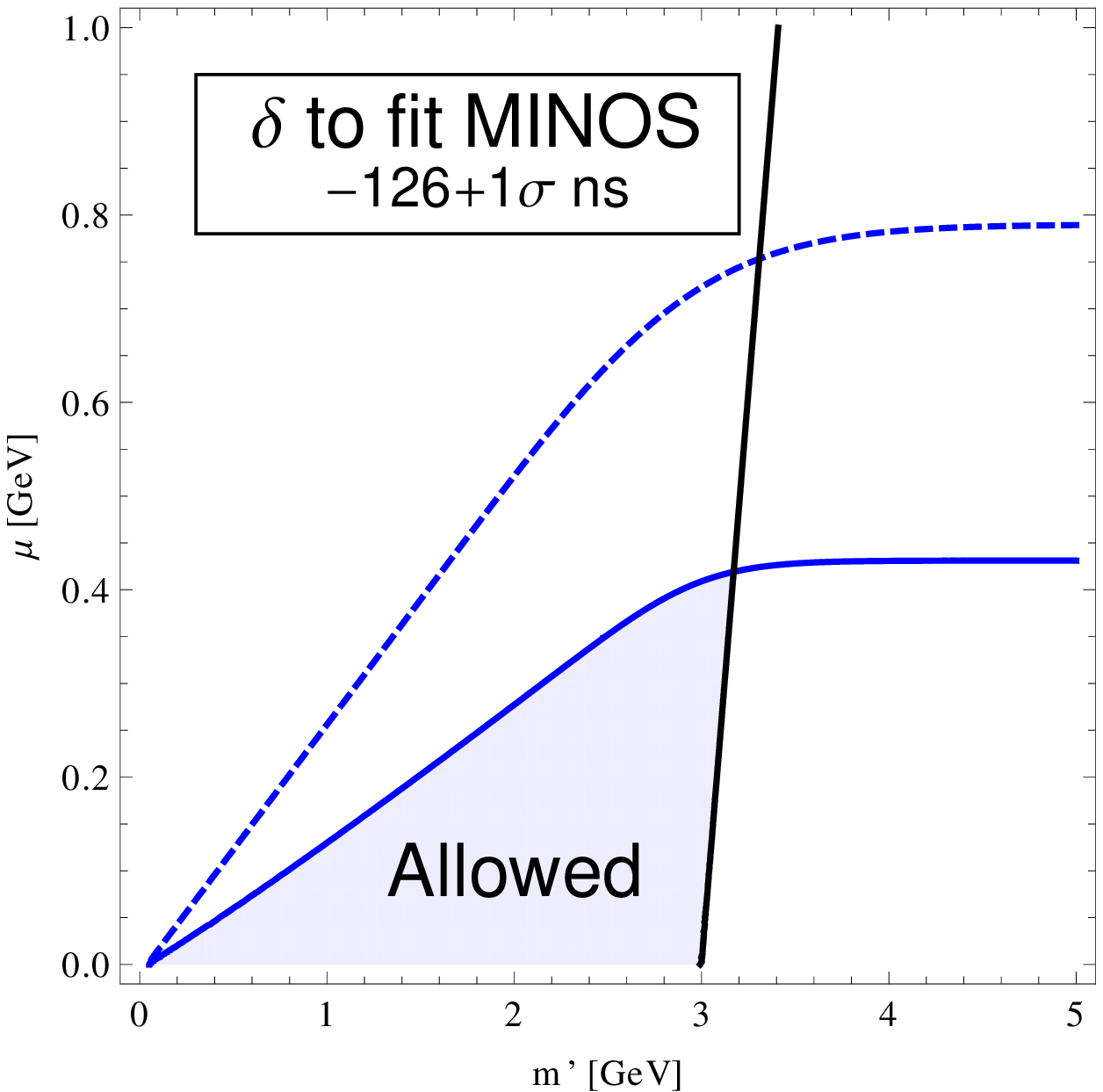, width=0.48\textwidth, angle=0}\label{fig:SN_MINOS_Tanh2a}} \hfill
\subfloat[]
{\epsfig{file=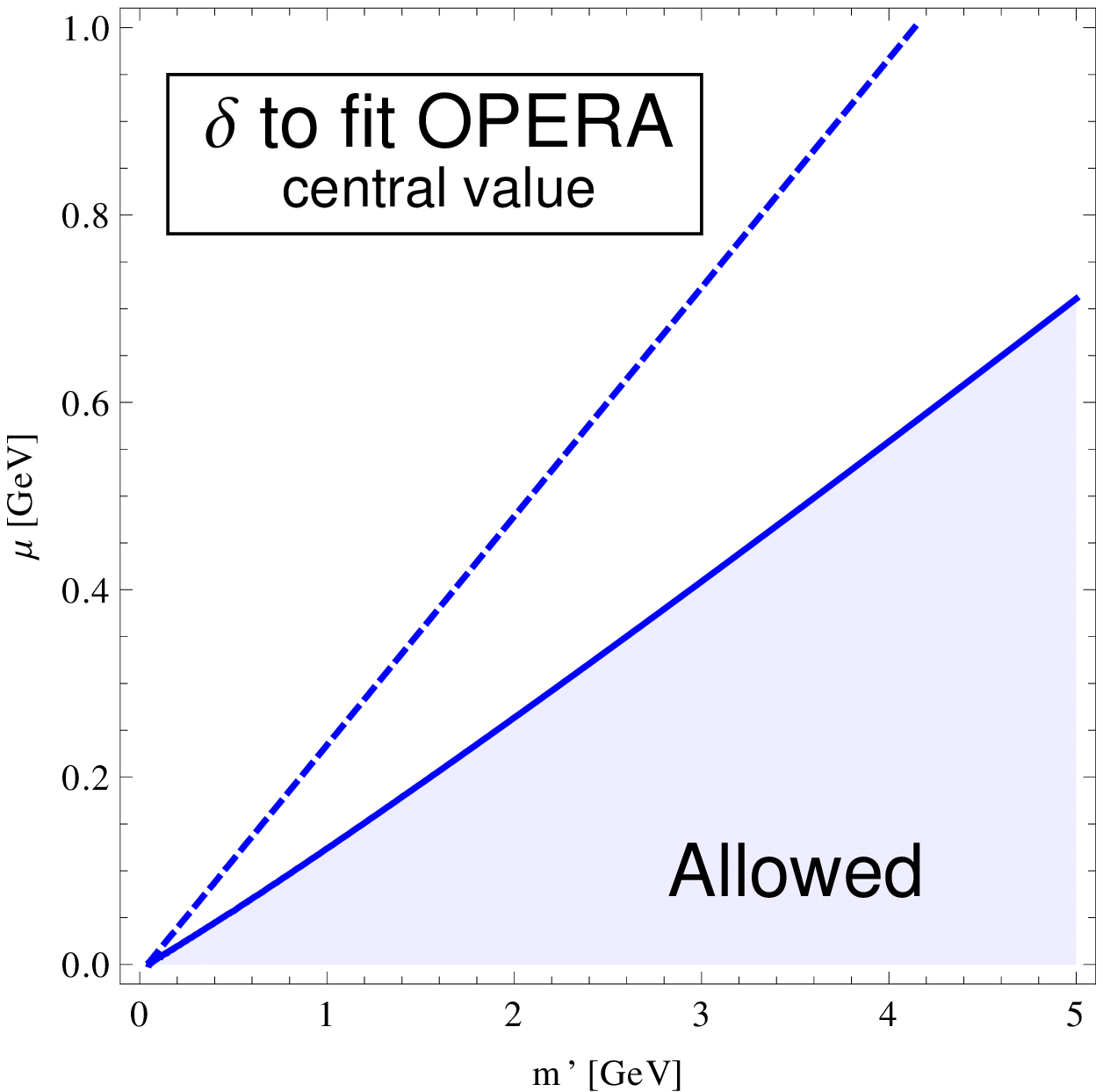, width=0.48\textwidth, angle=0}\label{fig:SN_MINOS_Tanh3}}
\caption{\footnotesize Bounds for the threshold $m^\prime$ and the spread $\mu$ of the step function after fitting the coefficient $\delta$ with MINOS and OPERA preferred regions.}
\label{fig:fit}
\end{center}
\end{figure}

The situation is much improved in the case of a step function (\ref{eq:step}): in fact, the energy ranges of supernova and MINOS 
neutrinos are far enough that a transition between a luminal propagation at low energies and a super-luminal propagation at MINOS or 
OPERA energies can be accommodated.
Fig.~\ref{fig:SN_MINOS_Tanh1} shows that for a transition at energies of $1$ GeV with a spread of less than $100$ MeV, both MINOS 
and OPERA results are fully compatible with high energy and supernova observations. Fig.~\ref{fig:SN_MINOS_Tanh1a}, on the other 
hand, shows that increasing the transition energy at above the average energy of MINOS neutrinos, only OPERA results can be made 
compatible with supernova and high energy neutrino bounds. This is still acceptable due to the low significance of the MINOS result.
If we fit the value of the parameter $\delta$ to reproduce either MINOS or OPERA results, we can obtain a more detailed information 
about the two mass scales in this scenario, namely the position of the transition $m'$ and the spread $\mu$.
As the central value of MINOS is excluded by the bound from high energy neutrinos, we fit a value at 1$\sigma$ from the central value, 
namely $\delta_t = -126 + 96$ ns: the allowed region is shown in Fig.~\ref{fig:SN_MINOS_Tanh2a}.
An analogous plot can be obtained fitting $\delta$ with the central value of OPERA, shown in Fig.~\ref{fig:SN_MINOS_Tanh3}.
These results show that the step function parametrisation has the potential to fit both MINOS and OPERA results without contradicting 
the bounds from the supernova SN1987a.
However, in order to establish the validity of this scenario, a detailed simulation of the three neutrino data sets is necessary to take into 
account the non trivial energy spectra of the supernova neutrinos and MINOS/OPERA beams.
We plan to perform such an analysis once the detailed data from OPERA will be publicly available.

\subsection{Comment on bounds from \v{C}erenkov-like emission}

In Ref.~\cite{Cohen:2011hx}, which appeared after our paper, it has been pointed out that super-luminal neutrinos, with a constant velocity matching the OPERA result, should lose energy in the form of electron pairs for energies above $140$ MeV. An explicit calculation of the energy loss along the distance between CERN and the OPERA detector shows that all neutrinos with energy above $12.5$ GeV should lose most of their energy before reaching Gran Sasso, therefore depleting the beam of all the high energy neutrinos that give rise to the events detected by OPERA. If this argument held, the detection of neutrinos by OPERA would contradict their super-luminality~\footnote{Bounds on neutrino production at CERN based on similar arguments have been pointed out in ~\cite{Bi:2011nd,GonzalezMestres:2011jc}.}. However, the argument relies on a specific form of Lorentz violation due to the presence of a preferred frame~\cite{AmelinoCamelia:2011bz}, and the calculation is based on a modified form of the dispersion relation and not directly on the value of the velocity. 
Therefore, the calculation of the energy loss can be performed only in a specific model where the origin of the Lorentz violation is specified and the link between LV and the velocity of neutrinos is expressed.
In this sense, the argument is not general and cannot be used to rule out a super-luminal interpretation of the OPERA measurement.

As an example where the \v{C}erenkov emission argument may not apply, we can quote the case of shortcuts via extra dimensions~\cite{Csaki:2000dm}, which is similar to the toy model described in this paper. In this case, based on a curved extra dimensional space-time where the other standard model particles are localised on a 4 dimensional Minkowski sub-manifold, neutrinos are assumed to propagate in the bulk of the extra dimension and, due to the curvature, they bounce back and forth on the standard model brane.
In the massless limit, their motion proceeds at the speed of light along the geodesics, therefore in the extra dimensional space-time there is no super-luminal propagation and the laws of relativity are not violated. However, the effect of the curvature can be such that the distance covered by the neutrino in the bulk is smaller than its projection on the standard model brane. Any experiment made of ordinary matter will measure the longer distance along the brane and therefore detect an apparent super-luminality. \v{C}erenkov emission does not take place because the true velocity of the neutrino is always lower that the speed of light. 
On the other hand, in a realistic model, neutrinos may lose energy when crossing the brane: however this problem is related not to the velocity but to the effective interactions of the extra dimensional neutrinos and therefore it can only be addressed once a full model is specified.

Even though the \v{C}erenkov-like energy loss is not a generic feature of super-luminal neutrinos and it can be calculated only once a specific model or class of models is specified, in the rest of this section we will follow the argument of Ref.~\cite{Cohen:2011hx} and estimate the effect of the energy dependence of the velocity on the \v{C}erenkov emission.
If we assume that the power law velocity in Eq.~\ref{eq:vel} can be derived from a modified dispersion relation with a preferred frame, the dispersion relation should have the form
\beq
E^2-\vec{p}^2 - \frac{1}{\alpha+1} \frac{E^{\alpha+2}}{M^\alpha}=0\,;
\eeq
with $v = \partial E/\partial p$.
Therefore, neutrinos have an effective energy-dependent mass
\beq
\delta m^2 (E) = \frac{1}{\alpha+1} \frac{E^{\alpha+2}}{M^\alpha}\,.
\eeq
Assuming that the \v{C}erenkov emission does take place, from the expression of the effective mass we can estimate the energy loss rate (Eq. 3 in Ref.~\cite{Cohen:2011hx}):
\beq
\frac{dE}{dx} = - \frac{25}{448} \frac{G_F^2}{192 \pi^3} (\delta m^2)^3\,.
\eeq
Integrating over the baseline of OPERA $L$, and using the relation $\delta_{\rm OPERA} =5 \times 10^{-5} = \left(\frac{\langle E \rangle}{M}\right)^\alpha$, where the average energy of detected neutrinos is $\langle E \rangle \sim30$ GeV, the energy $E_T$ above which the neutrinos lose their energy is given by the relation:
\beq
E_T^{5+3 \alpha} = \frac{(1+\alpha)^3}{5+3 \alpha} \frac{\langle E \rangle^3}{\frac{25}{448} \frac{G_F^2}{192 \pi^3} \delta_{\rm OPERA}^3 L}\,.
\eeq
For $\alpha = 0$, we recover the result in Ref~\cite{Cohen:2011hx}; by numerical study of the formula, we found that $E_T \gtrsim 30$ GeV for $\alpha \gtrsim 5$.
Thus, while large values of $\alpha$ would resolve the inconsistency of OPERA in the case of preferred frame Lorentz violation, this case is excluded both by the Fermilab bound and by the absence of bunch distortion, as pointed out in this paper.
As a concluding remark, we repeated the calculation in the case of an hyperbolic tangent step function: the result for the threshold energy is very close to $12.5$ GeV in the interesting parameter region, due to the fact that the velocity is almost energy-independent in the vicinity of the OPERA energies.
Thus, in the class of models where \v{C}erenkov emission takes place, an energy dependence in the velocity does not allow to evade the bounds from the energy loss of OPERA neutrinos.

\section{Conclusions}
\label{sec:concl}
\setcounter{equation}{0}
\setcounter{footnote}{0}

We presented a combined analysis of possible LV effects using the available data from SN1987a and 
the MINOS and OPERA neutrino velocity test. 
The MINOS collaboration reported a hint of super-luminal propagation for muonic neutrinos of a few GeV energy, even 
though the result is not statistically significant being compatible with the speed of light at $1.4 \sigma$ (summing linearly the systematic and statistic errors). The OPERA 
collaboration reported instead a more precise result which corresponds to a $6 \sigma$ effect for super-luminal 
propagation for muonic neutrinos, thus confirming the MINOS results. We studied the possible bounds on a general LV term in the velocity that depends on non integer 
powers of the energy, which are naturally generated in the context of conformal neutrinos.
We considered the distortion in shape of the bunch of neutrinos from MINOS at the far detector together with the time of 
flight measurement: in fact, the two effects are correlated to each other via the energy dependence of the LV term in the velocity.
We found a tension between the MINOS preferred region and the supernova bounds, coming from the fact that the 
parameter region with large energy dependence, which would be favoured by supernova data, is disfavoured by MINOS 
due to the absence of a shape distortion in the data. This effect has not been taken into account in previous studies.
The significance of the MINOS result alone does not allow to draw any solid conclusion, however the very recent results from OPERA will certainly help clarify the situation in the future.
In particular, the OPERA collaboration reported that they do not observe a marked energy dependence of the effect, thus supporting our conclusions.
We also tested other forms of energy dependence, and found that the most favourable one is close to a step function  
(hyperbolic tangent). This form allows to have velocity very close to the speed of light at low energy, thus evading 
bounds from supernova neutrinos, and an almost energy independent deviation at large energies, which accommodates the preferred region by OPERA and the absence of bunch distortion in MINOS.
These results point to new and partially unexpected behaviour which indicates a super-luminal time shift 
for GeV energy neutrinos with no or small bunch shape distortion due to the energy spread.
Neutrinos might therefore not only serve as a probe for physics beyond the Standard Model and cosmology, but also 
for the understanding of the foundations of space-time.

%



\end{document}